\documentclass[twoside,a4paper,11pt]{sea9}
\usepackage{graphicx}
\usepackage{hyperref}
\usepackage{movie15}
\pdfoutput=1
\begin{document}
\pagenumbering{arabic}
\pagestyle{myheadings}
\thispagestyle{empty}
{\flushleft\includegraphics[width=\textwidth,bb=58 650 590 680]{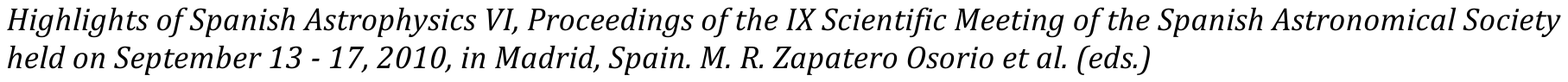}}
\vspace*{0.2cm}
\begin{flushleft}
{\bf {\LARGE
%
Science with the Galactic O-Star Spectroscopic Survey (GOSSS) :  The relationship between DIBs, ISM, and extinction.
%
}\\
\vspace*{1cm}
%
Miguel Penad\'es Ordaz$^{1}$,
Jes\'us Ma{\'\i}z Apell\'aniz$^{1}$, 
Alfredo Sota Ballano$^{1}$,
Emilio J. Alfaro$^{1}$,
Nolan R. Walborn$^{2}$,
Rodolfo H. Barb\'a$^{3}$,
Nidia I. Morrell$^{4}$,
Julia I. Arias$^{3}$,
and
Roberto C. Gamen$^{5}$
%
}\\
\vspace*{0.5cm}
%
\footnote{ \href{http://www.iaa.es}{www.iaa.es}}Instituto de Astrof{\'\i}sica de Andaluc{\'\i}a IAA-CSIC (Granada-Spain)\\
\footnote{ \href{http://www.stsci.edu}{www.stsci.edu}}Space Telescope Science Institute (Baltimore-MD-USA)\\
\footnote{ \href{http://www.userena.cl}{www.userena.cl}}Universidad de La Serena (La Serena-Chile)\\
\footnote{ \href{http://www.lco.cl}{www.lco.cl}}Las Campanas Observatory (La Serena-Chile)\\
\footnote{ \href{http://www.fcaglp.unlp.edu.ar/$\sim$gladys/ialp}{www.fcaglp.unlp.edu.ar/$\sim$gladys/ialp}}Instituto de Astrof{\'\i}sica de La Plata (CONICET, UNLP), (La Plata-Argentina)\\
%
\end{flushleft}
%
\markboth{
Science with GOSSS :  The relationship between DIBs, ISM, and extinction.
}{ 
%
Miguel Penad\'es Ordaz et al.
%
}
\thispagestyle{empty}
\vspace*{0.4cm}
\begin{minipage}[l]{0.09\textwidth}
\ 
\end{minipage}
\begin{minipage}[r]{0.9\textwidth}
\vspace{1cm}
\section*{Abstract}{\small
%
In this poster we show our preliminary analysis of DIBs (Diffuse Interstellar Bands) and other interstellar absorption lines with the purpose of understanding their origin and 
their relationship with extinction. We use the biggest Galactic O-star blue-violet spectroscopic sample ever (GOSSS, see contribution by Ma{\'\i}z Apell\'aniz). 
This sample allows a new insight on this topic because of the adequacy of O-star spectra, the sample number (700 and increasing, 400 used here), and their distribution in the MW 
disk. We confirm the high correlation coefficients between different DIBs and $E(B-V)$, though the detailed behavior of each case shows small differences.
We also detect a moderately low correlation coefficient between Ca\,{\sc ii} $\lambda$3934 (Ca K) and $E(B-V)$ 
with a peculiar spatial distribution that we ascribe to the relationship between line saturation and velocity profiles for Ca\,{\sc ii} $\lambda$3934. 
%
\normalsize}
\end{minipage}
%
%
%
\section{Data description and processing\label{intro}}

We used $\sim$900 spectra corresponding to $\sim$400 O stars (plus some B and WR stars) from the GOSSS
survey (see talk by Jes\'us Ma{\'\i}z Apell\'aniz~\cite{Maiz Apellaniz 2010} in this meeting) and reduced using the 
 
\begin{figure}
\center
\includegraphics[scale=0.445]{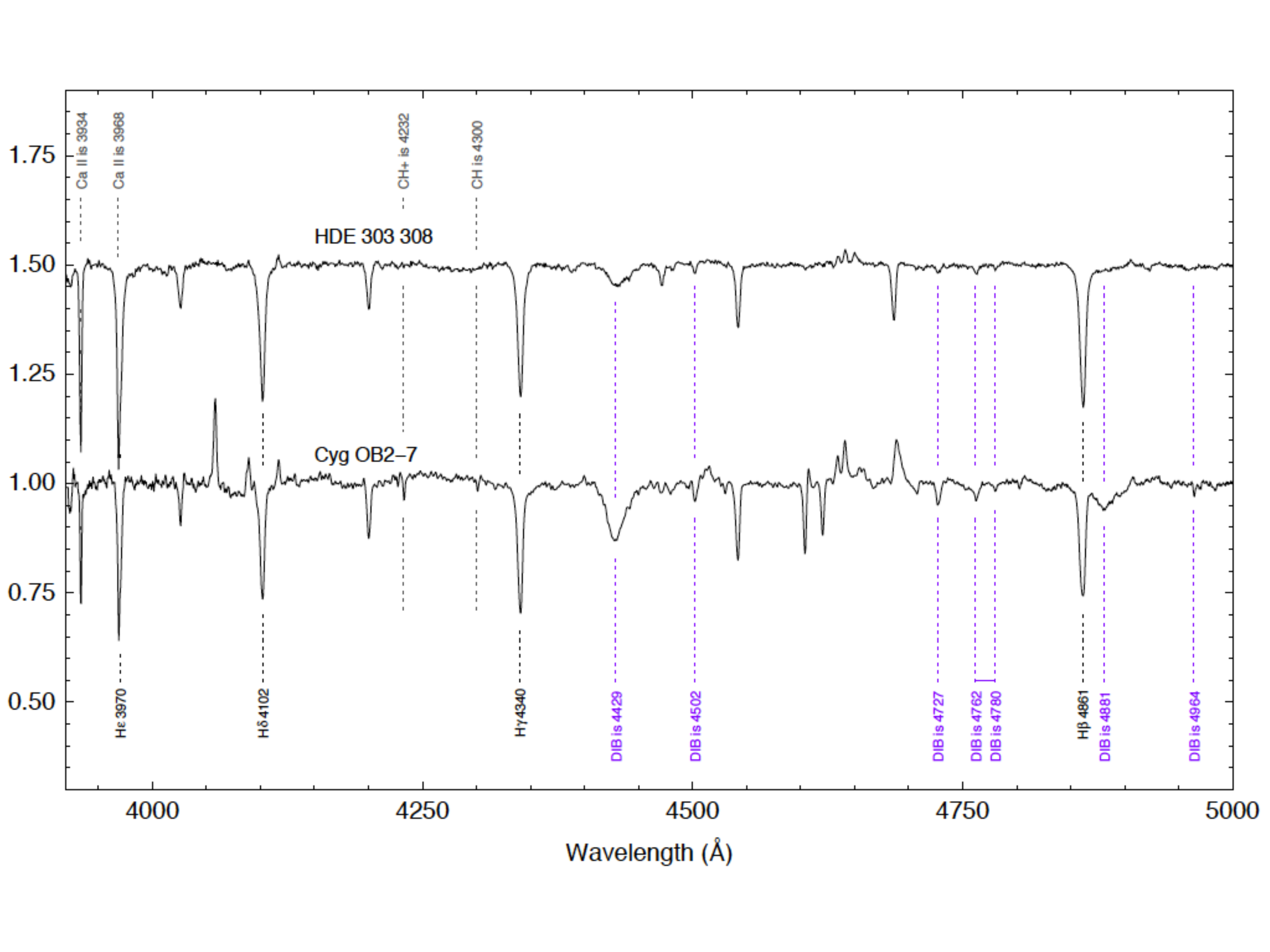} 
\caption{\label{fig1} Sample blue-violet spectra for two GOSSS O stars. The interstellar lines
and DIB bands studied here as well as some other relevant lines and bands are
indicated. Note the high $EW$ of the Ca II lines and the moderate $EW$ of the DIBs in
HDE 303 308 (in Carina) and compare it with the high $EW$ of the DIBs and moderate $EW$
of the Ca II lines in Cyg OB2-7 (in Cygnus).
}
\end{figure}

pipeline
developed by one of us (see poster by Alfredo Sota~\cite{Sota 2010} in this meeting). The spectra have high S/N
(200 or better) and cover the 3900-5100 \AA ~range with a spectra resolution R$\sim$2500.
To measure the equivalent widths of the bands and lines of our study (Ca II $\lambda$3934, CH+ $\lambda$4232,
and DIBs $\lambda$$\lambda$4501, 4726, 4762, 4780, and 4964 \AA) we have used a visual IDL package based
program written by one of us (Jes\'us Ma{\'\i}z Apell\'aniz). The code allowed us to perform a fast and
reliable analysis using both numerical integration and gaussian fits of each absorption line.
Each fit was visually inspected and interactively adjusted to account

\begin{table}[ht] 
\caption{Correlation coefficients (top row in each cell) and noise-modified maximum
correlation coefficients (bottom row) for the quantities in this work.} 
\center
\includegraphics[scale=0.425]{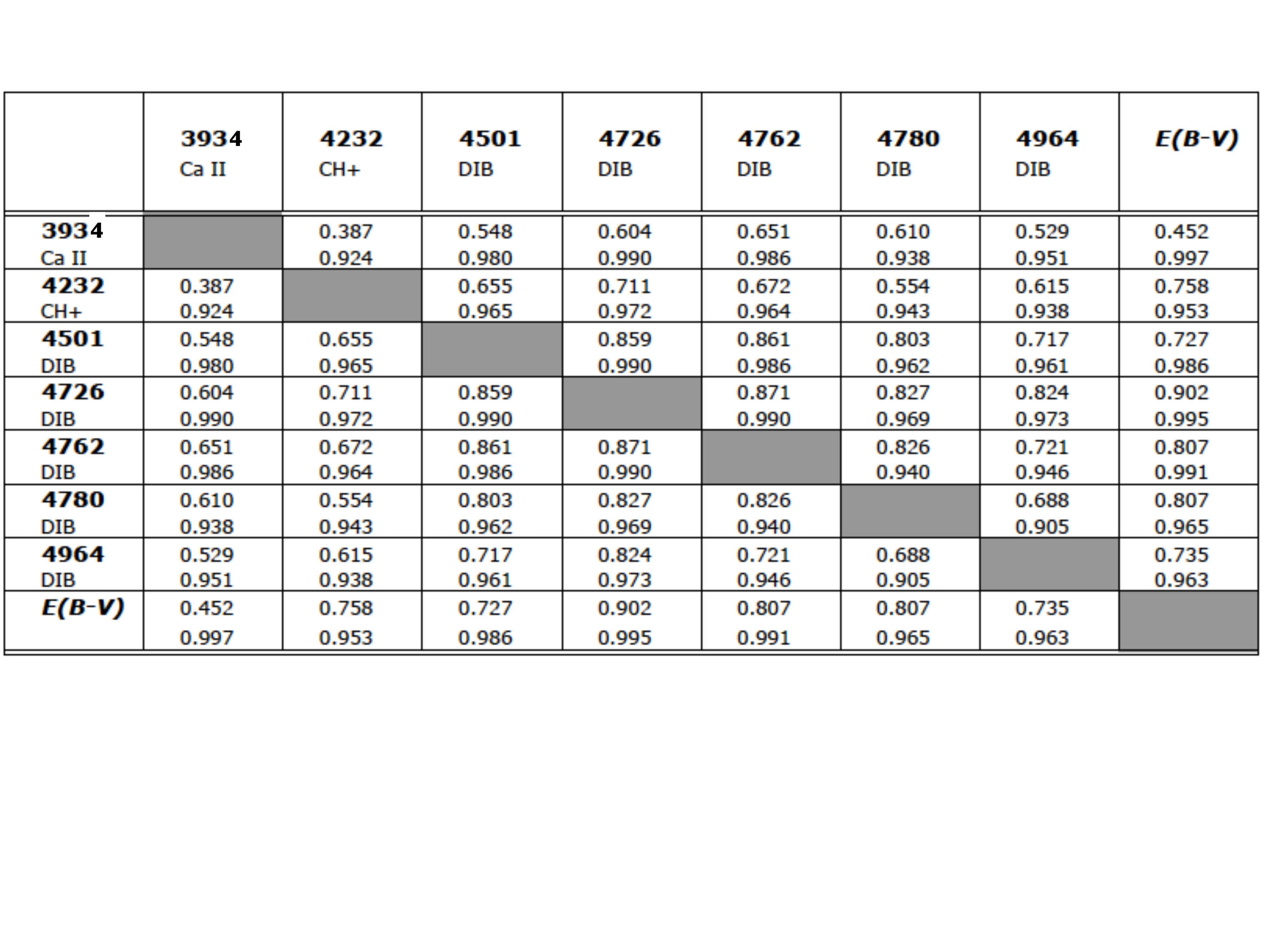} \\
\label{tab1} 
\end{table}
for contamination by
nearby lines, velocity shifts, and S/N effects. Some of the spectra were eliminated because of
quality-control issues related to those effects.
The color excesses $E(B-V)$ were derived from [a] the photometry in the Galactic O-Star Catalog
(Ma{\'\i}z Apell\'aniz et al. 2004~\cite{Maiz Apellaniz 2004b}, Sota et al. 2008~\cite{Sota 2008}), [b] the temperature-spectral type
relationships in Martins et al. (2005)~\cite{Martins 2005}, and [c] the magnitudes as a function of temperature
and luminosity class in the latest version of CHORIZOS (Ma{\'\i}z Apell\'aniz 2004~\cite{Maiz Apellaniz 2004a}).
All the equivalent widths were correlated with each other and with the color excess Table~\ref{tab1}.
We performed for each pair linear and parabolic fits. For the linear fits we calculated the
maximum possible correlation coefficient for two perfectly correlated variables affected by
random noise (the value is close to but not exactly 1.0).

\section{Results\label{results}}
DIBs are one of the older questions without answer in astronomy. More than 300 bands have been
discovered in nearly a century. The currently favored carriers are carbon-based, but their
specific nature is still highly debated (Herbig 1995~\cite{Herbig 1995}, Galazutdinov et al. 2000~\cite{Galazutdinov 2000}, Ka\'zmierczak et
al. 2010~\cite{Kazmierczak 2010}).
As we knew from previous literature results, DIBs are tightly correlated with extinction, but
also with a considerable dispersion and differences for each band. This has been once again
verified in this study. Other important corroboration is the variability of the correlation
coefficients, indicating that DIBs are likely to originate in a family of carriers instead of
a single one (Table~\ref{tab1}).

\begin{figure}
\center
\includegraphics[scale=0.35]{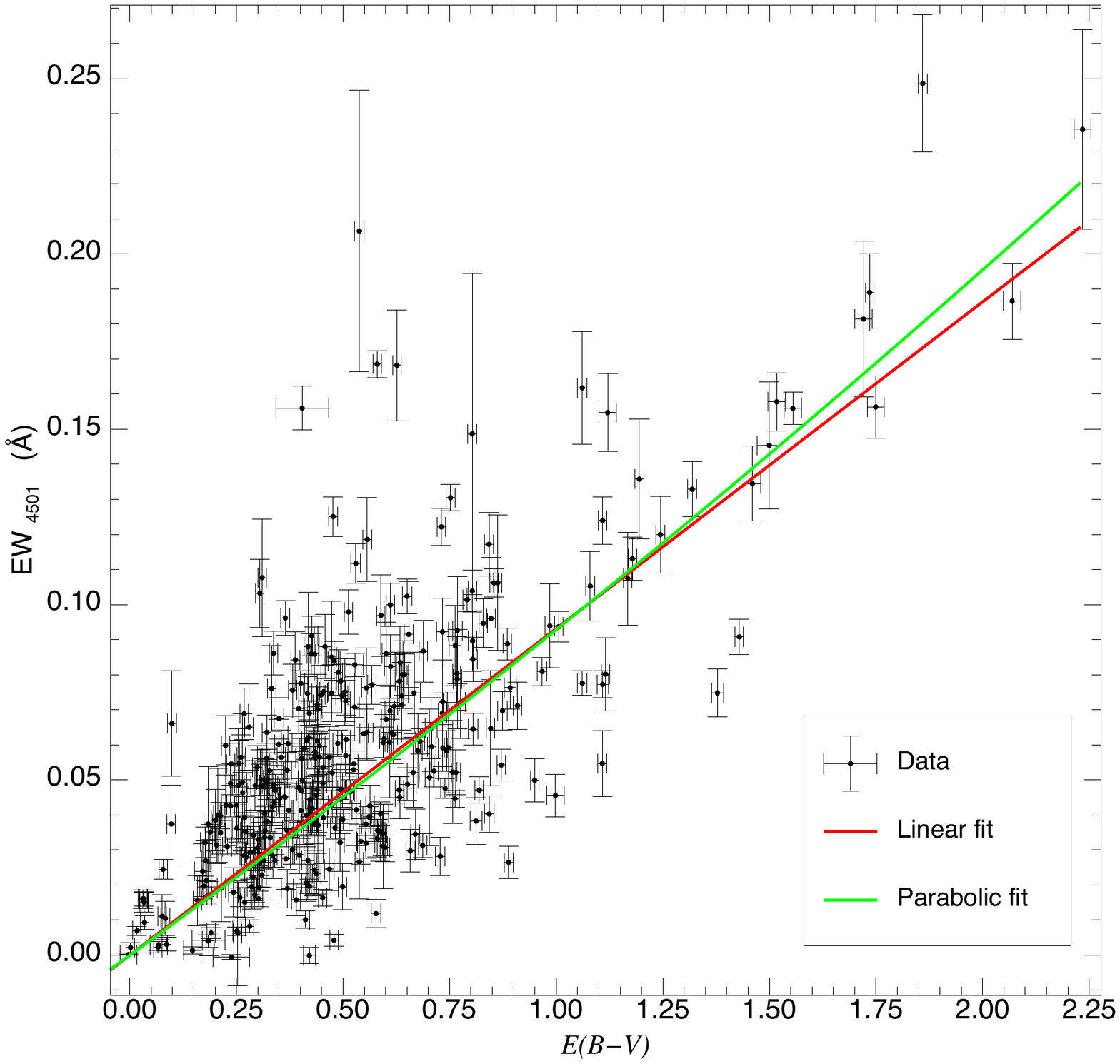} ~
\includegraphics[scale=0.35]{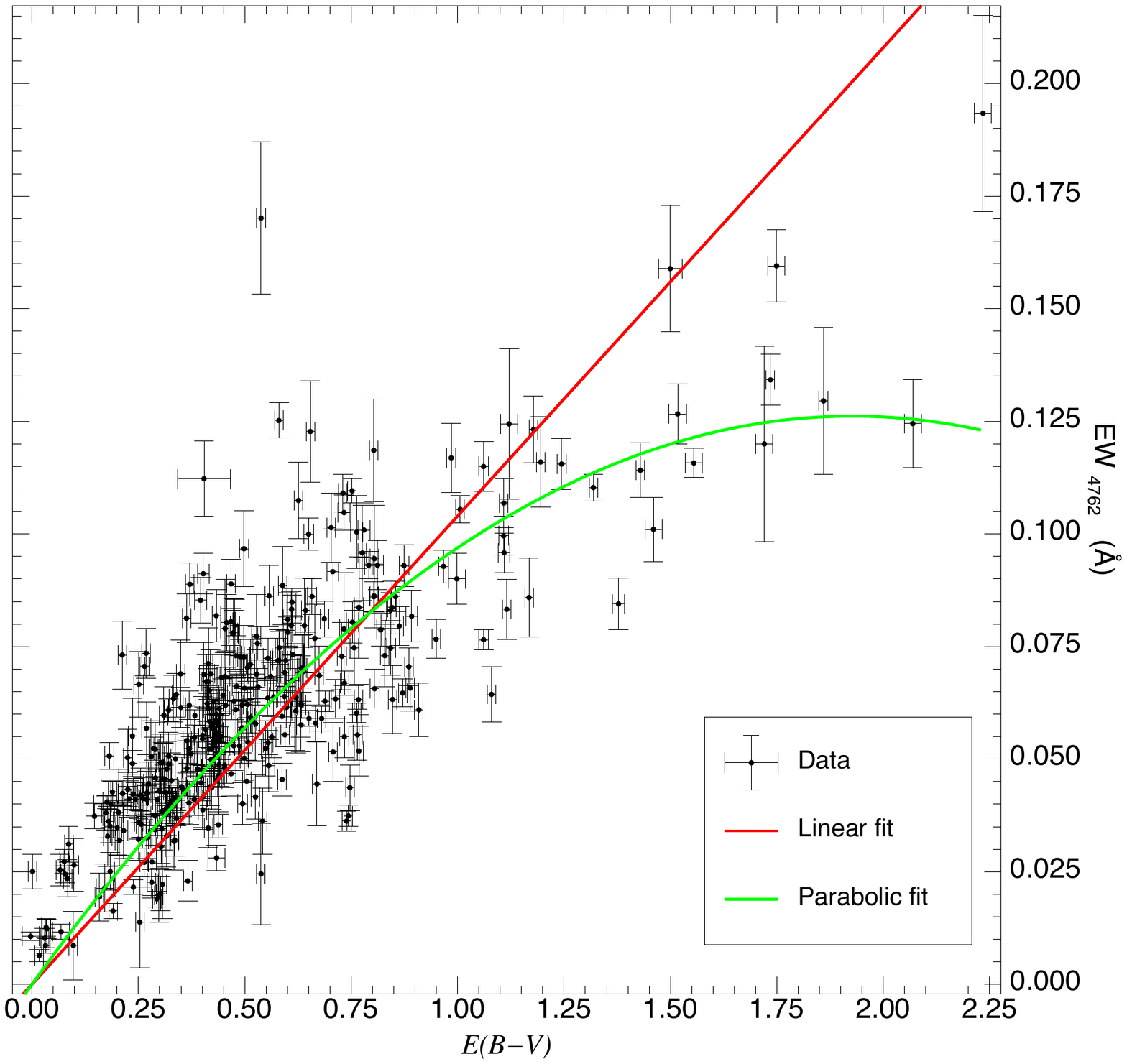} 
\caption{\label{fig2} Equivalent line widths plotted against color excess for the
two DIB bands centered near 4501 \AA ~(left) and 4762 \AA ~(right). We also show the linear
and parabolic fits performed. The better fit by a parabola for the 4762 \AA ~DIB band behavior
is possibly due to the skin effect (DIBs originating in the diffuse ISM and in the outer
layers of molecular clouds).
}
\end{figure}

It is believed that DIB carriers are present in diffuse clouds or the surface of dense ones as
opposed to dense cores. This could be the reason of the better fit by a parabola in Fig. ~\ref{fig2} (right) and
is known as the skin effect: at high extinctions the $EW$ of some DIBs depends on color excess
more weakly than at low extinctions. On the other hand, some DIBs do not show the effect in
our data (Fig. ~\ref{fig2}).

\begin{figure}
\includegraphics[scale=0.35]{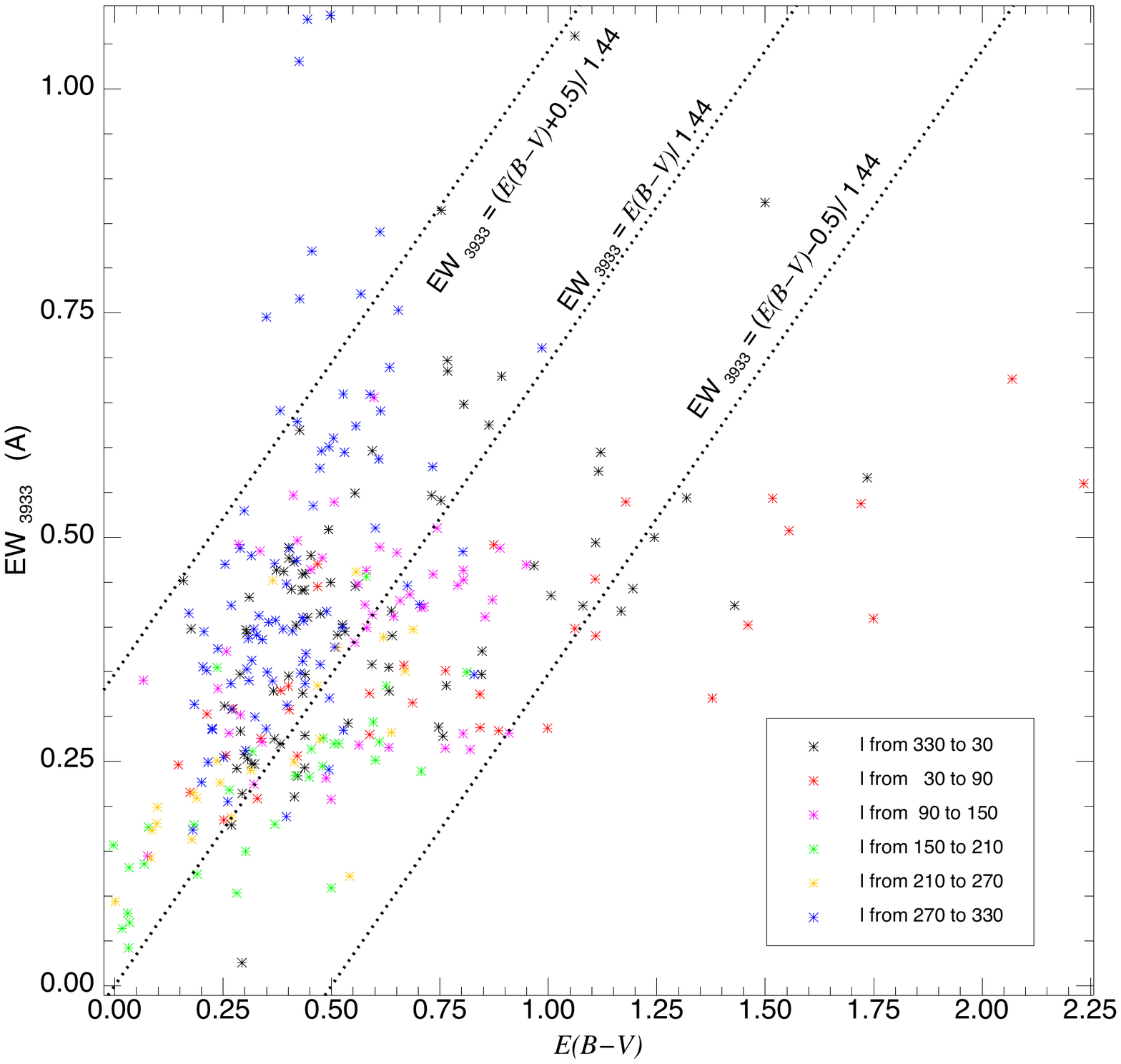} ~
\includegraphics[scale=0.35]{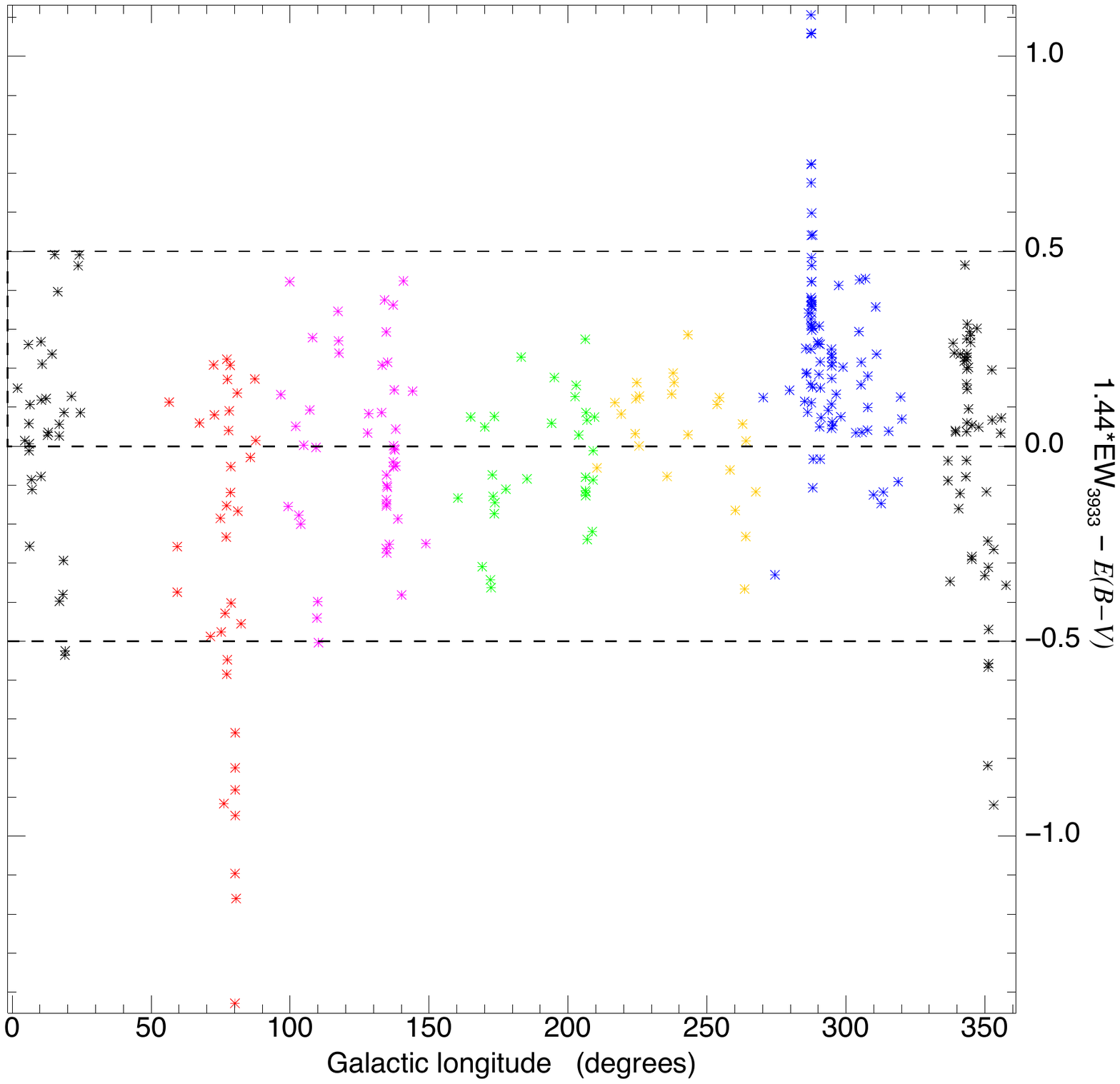} 
\caption{\label{fig3} Behavior of the Ca II $\lambda$3934 (K) interstellar absorption
as a function of $E(B-V)$ and Galactic longitude for the GOSSS sample. Note the different
behavior in the direction of, on the one hand, Carina (blue), and on the other hand,
Cygnus (red points) and, to some extent, Sagittarius (black). The 1.44 value is the slope
of the linear fit performed.
}
\end{figure}

The relationship of the Ca II $\lambda$3934 line with extinction Figs.~\ref{fig3} shows an interesting
behavior. For low values of $E(B-V)$, the $EW$ of the line is approximately proportional to the
reddening. At intermediate values, however, two branches form, one with large $E(B-V)$ and
intermediate $EW(3934)$ and another one with intermediate $E(B-V)$ and large $EW(3934)$.
Furthermore, the two branches are distinguished as regions in the sky, with the first one
concentrated in Cygnus (with most stars being Cyg OB2 members) and, to some extent,
Sagittarius, and the second one in Carina. The most plausible explanation is the existence of
cloud velocity structure (unresolved in our spectra). Ca II $\lambda$3934 can saturate in moderately
dense clouds, so the behavior of its equivalent width depends not only on the amount of
material present but also on whether it is distributed in one or more clouds of different
velocities. Fig. \ref{fig3} is consistent with Ca II being concentrated in a single cloud in
the direction of Cygnus and in several in the direction of Carina. Such velocity distribution
has been extensively studied for the case of the Carina Nebula Asssociation (Walborn et al.
2007~\cite{Walborn 2007} and references therein), where multiple velocity components are indeed present and caused
by the kinetic energy input from the massive stars there. On the other hand, Cygnus OB2 is
closer to us than the Carina Nebula and has no associated H II region, so the absorbing Ca II
cloud is likely to be a single and relatively unperturbed foreground object.


\section{Future \label{future}}
Our future plans include:
\begin{itemize}
 \item[1] Increase the sample of observed stars.
 \item[2] Analyze the rest of the DIB lines in the blue-violet spectra.
 \item[3] Obtain spectra in the rest of the visible spectrum to observe more DIBs.
 \item[4] Use high-resolution spectra to study the velocity structure of the Ca II $\lambda$3934 line and
weak DIBs.
 \item[5] Use CHORIZOS to derive $E(4405-5495)$ and $R_{5495}$, the monochromatic equivalents to $E(B-V)$ and
$R_{V}$, respectively, and study their correlation with the measured equivalent widths.
\end{itemize}

%
\small  
%
\section*{Acknowledgments}   
%
Support for this work was provided by [a] the Spanish Government Ministerio de Ciencia e Innovaci\'on through grants AYA2007-64052, the Ram\'on y Cajal Fellowship program, and FEDER funds; [b] the Junta de Andaluc{\'\i}a grant P08-TIC-4075; and [c] NASA through grants GO-10205, GO-10602, and GO-10898 from the Space Telescope Science Institute, which is operated by the Association of Universities for Research in Astronomy Inc., under NASA contract NAS 5-26555.This research has made extensive use of Aladin (Bonnarel et al., 2000), and of the SIMBAD database,
operated at CDS, Strasbourg, France.

%

%
\end{document}